\pdfoutput=1

\documentclass{appolb}

\usepackage{amssymb}
\usepackage{hyperref}
\usepackage{graphicx}
\usepackage{amsmath}
\usepackage[numbers,sectionbib,sort&compress]{natbib}

\begin{document}
\title{Off-shell generalized parton distributions of the pion%
\thanks{Talk presented by WB at {\em XXIX The Cracow Epiphany Conference}, Cracow, Poland, 16-19 January 2023}}

\author{Wojciech Broniowski$^{1,2}$\thanks{Wojciech.Broniowski@ifj.edu.pl},\\
Vanamali Shastry$^{1}$\thanks{vanamalishastry@gmail.com}, 
Enrique Ruiz Arriola$^3$\thanks{earriola@ugr.es}
\address{$^{1}$Institute of Physics, Jan Kochanowski University, PL-25406~Kielce, Poland}
\address{$^{2}$The H. Niewodnicza\'nski Institute of Nuclear Physics, \\ Polish Academy of Sciences, PL-31342~Cracow, Poland}
\address{$^{3}$Departamento de F\'isica At\'omica, Molecular y Nuclear \\ and Instituto Carlos I de F\'{\i}sica Te\'orica y Computacional,
                   Universidad de Granada, E-18071 Granada, Spain}
}

\maketitle

\abstract{
We analyze off-shell effects in the generalized parton distributions (GPDs) of the pion in the context of the Sullivan electroproduction process, as well as the corresponding half-off-shell electromagnetic and gravitational form factors. We illustrate our general results within a chiral quark model, where the off-shell
effects show up at a significant level, indicating their contribution to uncertainties in the extraction of the GPDs from the future experimental data.
}

\bigskip
\bigskip

\begin{figure}[t]
\centering
\hfill \includegraphics[angle=0,width=0.36 \textwidth]{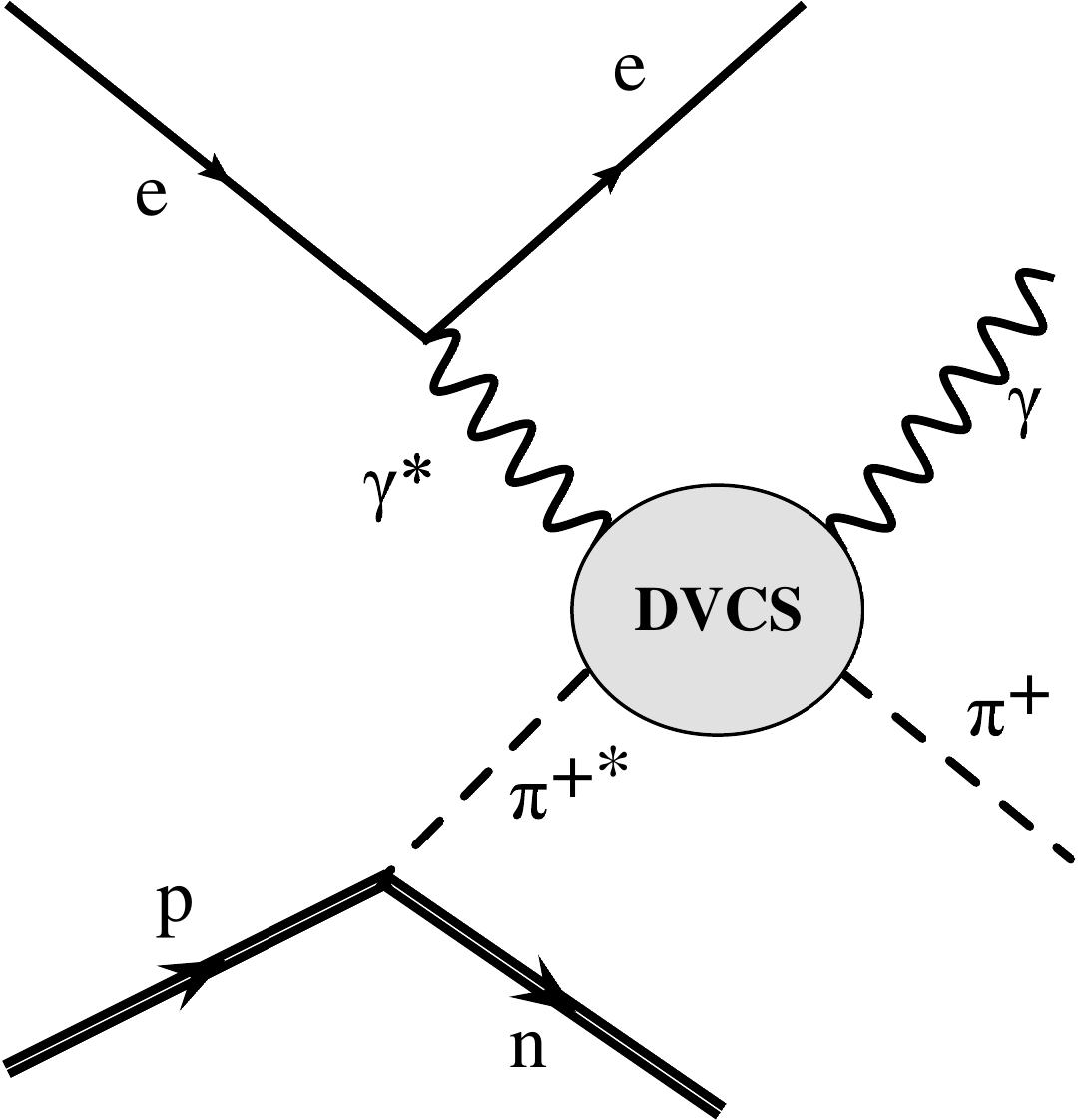} \hfill \hfill \includegraphics[angle=0,width=0.38 \textwidth]{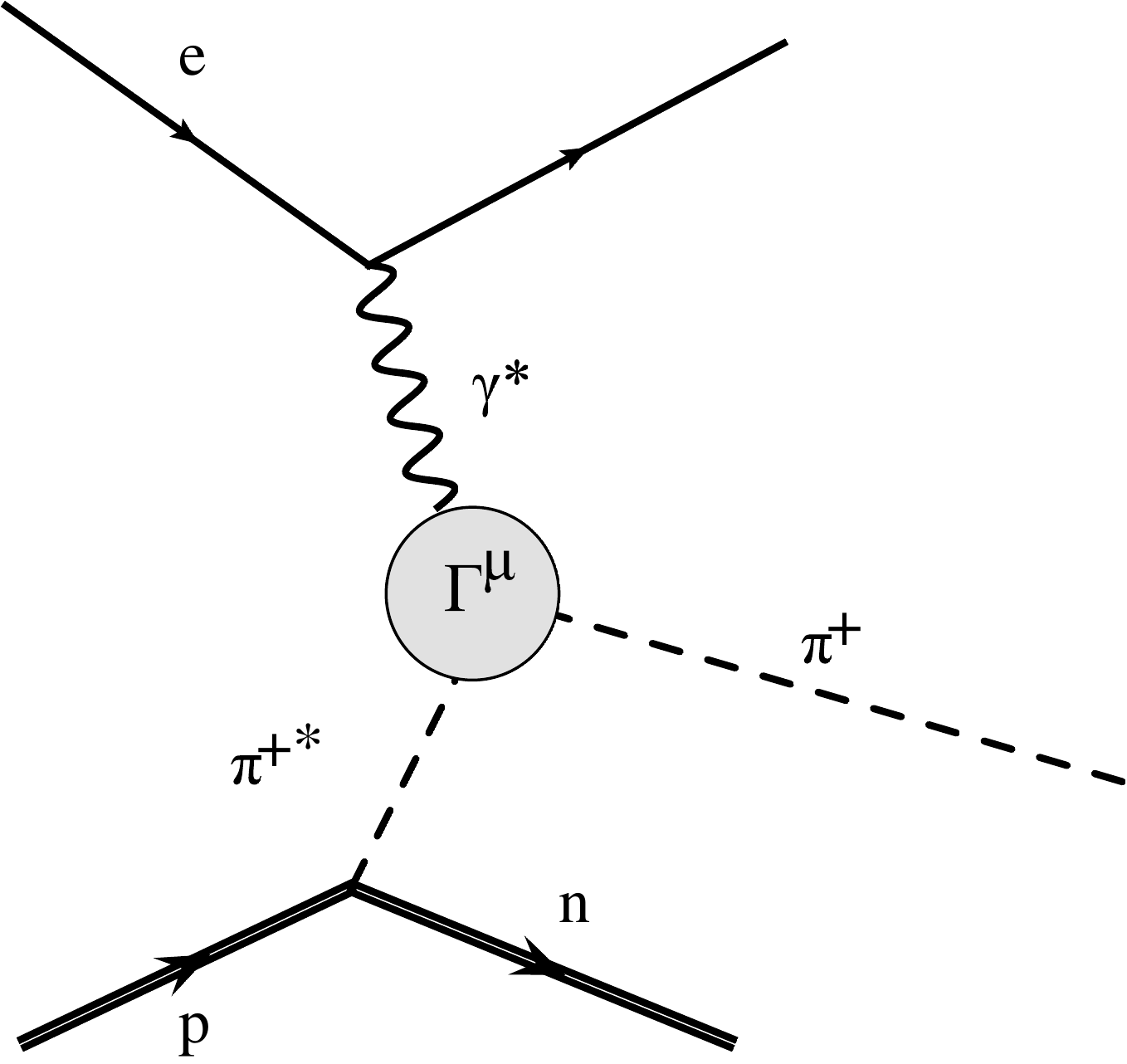} \hfill
\vspace{1mm}
\caption{Left: The Sullivan process for the pion electroproduction off the proton, involving the deeply virtual Compton scattering (DVCS) amplitude. 
Right: The corresponding Bethe-Heitler amplitude with the half-off-shell pion form factor $\Gamma^\mu$ (the contribution with the photon emission from the initial electron not shown).
Asterisks indicate off-shellness.
\label{fig:sul}}
\end{figure}

This talk is based on our recent paper~\cite{Broniowski:2022iip}.
In the wake of electron-ion colliders, one hopes for the accessibility of the pion's GPDs~\cite{Ji:1996nm,Radyushkin:1997ki,Diehl:2003ny} in future experiments~\cite{Aguilar:2019teb,Chavez:2021koz}, such as the Sullivan electroproduction process~\cite{PhysRevD.5.1732} shown in the left panel of Fig.~\ref{fig:sul}. As one of the pions entering the deeply virtual Compton scattering (DVCS) amplitude is virtual, one needs to consider the off-shell effects from the start. 

The history of off-shell effects is rather old~\cite{Ekstein:1960xkd},
and lies at the root of the inverse scattering problem in Quantum
Mechanics and extends to Quantum Field Theory. In short, the only
physical quantity is the $S$-matrix, and the Green's functions used to
construct it are not themselves observable except at sufficiently long
times where only on-shell configurations
remain~\cite{tHooft:1973wag}. For instance, a direct calculation of
Green functions in Effective Field Theory does not necessarily
guarantee off-shell finiteness from on shell renormalization
conditions and suitable field redefinitions may be requested to ensure
off-shell renormalizability~\cite{Appelquist:1980ae}.  This brings in
the freedom of the field reparametrization invariance. Only in the
1990's, within the context of a possible experimental program to
determine off-shell effects in hadronic form factors, was it realized
and emphasized that the off-shell effects cannot be measured as a
physical observable even at the lowest orders in the chiral
perturbation theory for the case of the pion (see,
e.g., \cite{Fearing:1999im,Scherer:2000hh} and references
therein). They are model or scheme dependent, in particular, they
depend on the chosen parameterization of the pion field. The point
emphasized in~\cite{Broniowski:2022iip} is that {\em because} they are
computed in models or employed in experimental analyses, one should
consider symmetry constraints in the form of the Ward-Takahashi
identities, which are valid regardless of the scheme or choice of the
interpolating field.

If, however, one were able to evaluate the full cross section $ep \to
en\pi^+$ in a model (or simulate it on the lattice), one could compare
it directly to the experiment. There, the pion would not be
approximated with a pole term or a model propagator, but all the
hadronic (quark and gluon) processes would contribute, whereby the
off-shell effects would not appear. This utopia, however, is not only
currently impossible, but also not desired, as theoretically we wish
to have components (building blocks, here for the $p \to \pi^{+\ast}
n$ and $\gamma^\ast \pi^{+\ast} \to \gamma \pi^+$ processes) of the
amplitude, such as DVCS, which upon factorization (always to be
proved) enters also other physical processes. Therefore one is bound
to an evaluation of the building blocks, where in the case where we
apply intermediate hadronic states, the off-shellness needs to be
tackled with.

\begin{figure}[t]
\centering
\includegraphics[angle=0,width=0.48 \textwidth]{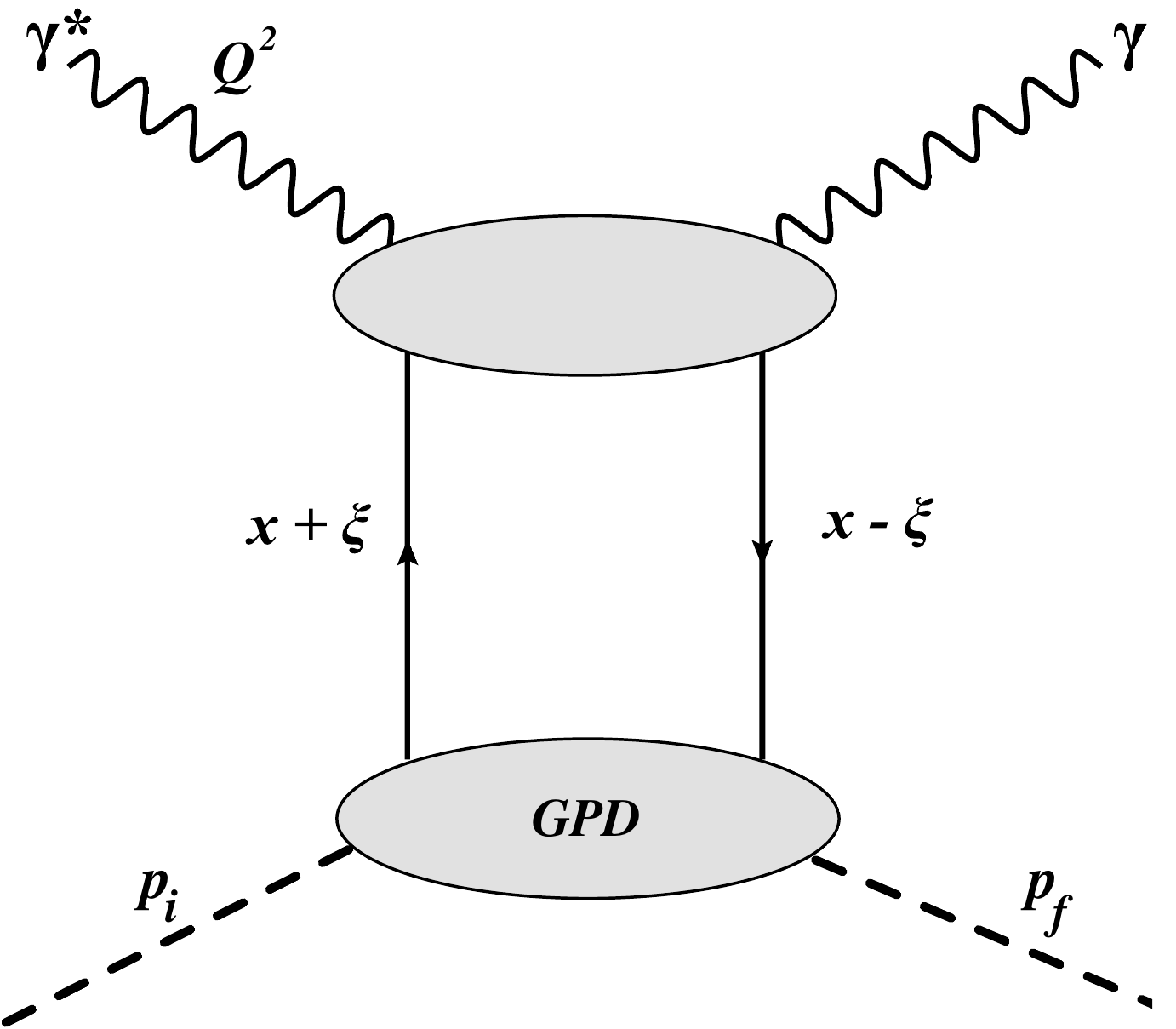} 
\vspace{1mm}
\caption{Factorization diagram for the DVCS amplitude into the perturbative part (the upper blob) and the non-perturbative GPD. 
\label{fig:dvcs}}
\end{figure}

The DVCS amplitude is related via QCD factorization to the GPD, as indicated in Fig.~\ref{fig:dvcs}. The off-shell quark GPDs of the pion are defined~\cite{Diehl:2003ny} as
\begin{eqnarray}
&&\hspace{-7mm} \delta_{ab}\delta_{\alpha\beta}{H}^{0}(x,\xi,t,p_i^2,p_f^2)+i\epsilon^{abc}\tau^c_{\alpha\beta}{H}^{1}(x,\xi,t,p_i^2,p_f^2) = \nonumber \\
&&\hspace{-5mm} \left .  \int \! \frac{d z^-}{4\pi} e^{i x \, P^+ z^-} \langle\pi^b(p_f)|\overline{\psi}_\alpha(-\tfrac{z}{2}) \gamma^+ \psi_\beta(\tfrac{z}{2})|\pi^a(p_i)\rangle \right |_{\substack{z^+=0\\z^\perp=0}}, \label{eq:Hdef}
\end{eqnarray}
and similarly for the gluon GPD, ${H}^{g}(x,\xi,t,p_i^2,p_f^2)$. Here $\psi$ indicates the quark field, $\alpha$ and $\beta$ are the quark flavors, $a$, $b$, and $c$ are the isospin indices, and the color summation is understood.  The subscripts $0$ and $1$ indicate the isosinglet and isotriplet quark
GPDs. The light-cone coordinates are defined with the convention $v^\pm= (v^0 \pm v^3)/\sqrt{2}$. In
the assumed light-cone gauge the Wilson links do not appear. The notation for the kinematics is
\begin{eqnarray}
P^\mu=\tfrac{1}{2} (p_f^\mu + p_i^\mu), \;\; q^\mu=p_f^\mu - p_i^\mu,  \;\;  t=q^2, \;\; \xi=-\frac{q^+}{2P^+}. \label{eq:not}
\end{eqnarray}
In the on-shell case $p_f^2=p_i^2=m_\pi^2$. In the partonic interpretation, $(x+\xi)P^+$ is the longitudinal momentum
of the struck parton (cf.~Fig.~\ref{fig:dvcs}). Importantly, the objects $H^{0,1,g}$ depend on the scale; we shall model their QCD evolution with the standard DGLAP-ERBL equations~\cite{Muller:1994ses,Diehl:2003ny}.  

For $p_f^2=p_i^2 $ time-reversal (the crossing symmetry) makes the GPDs of Eq.~(\ref{eq:Hdef}) even functions of $\xi$.  This is no longer the case when $p_f^2 \neq p_i^2$, as e.g. in the Sullivan process of Fig.~\ref{fig:sul}.  In this general case
the $x$-moments of the GPDs contain also {\em odd} powers of the skewness parameter $\xi$,
\begin{eqnarray}
\int_{-1}^1 dx \, x^j H^s(x,\xi,t,p_i^2,p_f^2)=\sum_{i=0}^{j+1} A^{s}_{j,i}(t,p_i^2,p_f^2) \xi^i, \;\;\; s=0,1,g, \label{eq:poly}
\end{eqnarray}
where $A^{s}_{j,i}$ are the generalized off-shell form factors. The most relevant ones are the form factors related to the
(conserved) electromagnetic and energy-stress tensor currents, since they do not depend on the factorization
scale. They correspond to the lowest $x$-moments:
\begin{eqnarray}
\int_{-1}^1 dx \,  {H}^{1}=2(F - G\xi), \;\;\; \int_{-1}^1 dx \,  x[H^{0}+H^{g}]=\theta_2 - \theta_3\xi - \theta_1\xi^2,  \label{eq:poly2} 
\end{eqnarray}
where, $F$ and $G$ are the electromagnetic form factors, $\theta_{1,2,3}$ are the gravitational form factors, and the dependence of the form factors on $(t,p_i^2,p_f^2)$ is understood. 

In~\cite{Broniowski:2022iip} we have discussed in detail the constraints following from the Ward-Takahashi identities for the charge form factors~\cite{Naus:1989em,Rudy:1994qb} $F$ and $G$, and for the gravitational form factors $\theta_{2,3}$. With the current-algebra techniques \cite{Schnitzer:1967zzb,Raman:1971jg} one derives\footnote{The tacit assumption here is that the pion field satisfies the PCAC (partially conserved axial current) condition, but it does not have to be elementary, i.e., can possess structure.} that
\begin{eqnarray}
G(t,p_i^2,p_f^2) &=& \frac{(p_f^2\!-\!p_i^2)}{t} \left [  F(0,p_i^2,p_f^2) -  F(t,p_i^2,p_f^2)\right ], \nonumber \\
\theta_3(t,p_i^2,p_f^2) &=& \frac{(p_f^2\!-\!p_i^2)}{t} \left [  \theta_2(0,p_i^2,p_f^2) -  \theta_2(t,p_i^2,p_f^2)\right ], \label{eq:t2ft3}
\end{eqnarray}
We remark that the form factor $\theta_1$ corresponds to a transverse tensor, hence is not constrained by the current conservation. In the chiral limit and on-shell, a low-energy theorem states that $\theta_1(0,0,0)= \theta_2(0,0,0)$~\cite{Donoghue:1991qv}.
The off-shell electromagnetic form factor $G(t,p^2,m_\pi^2)/p^2$ was recently examined phenomenologically in~\cite{Choi:2019nvk}.

\begin{figure}[t]
\centering
\includegraphics[angle=0,width=0.487 \textwidth]{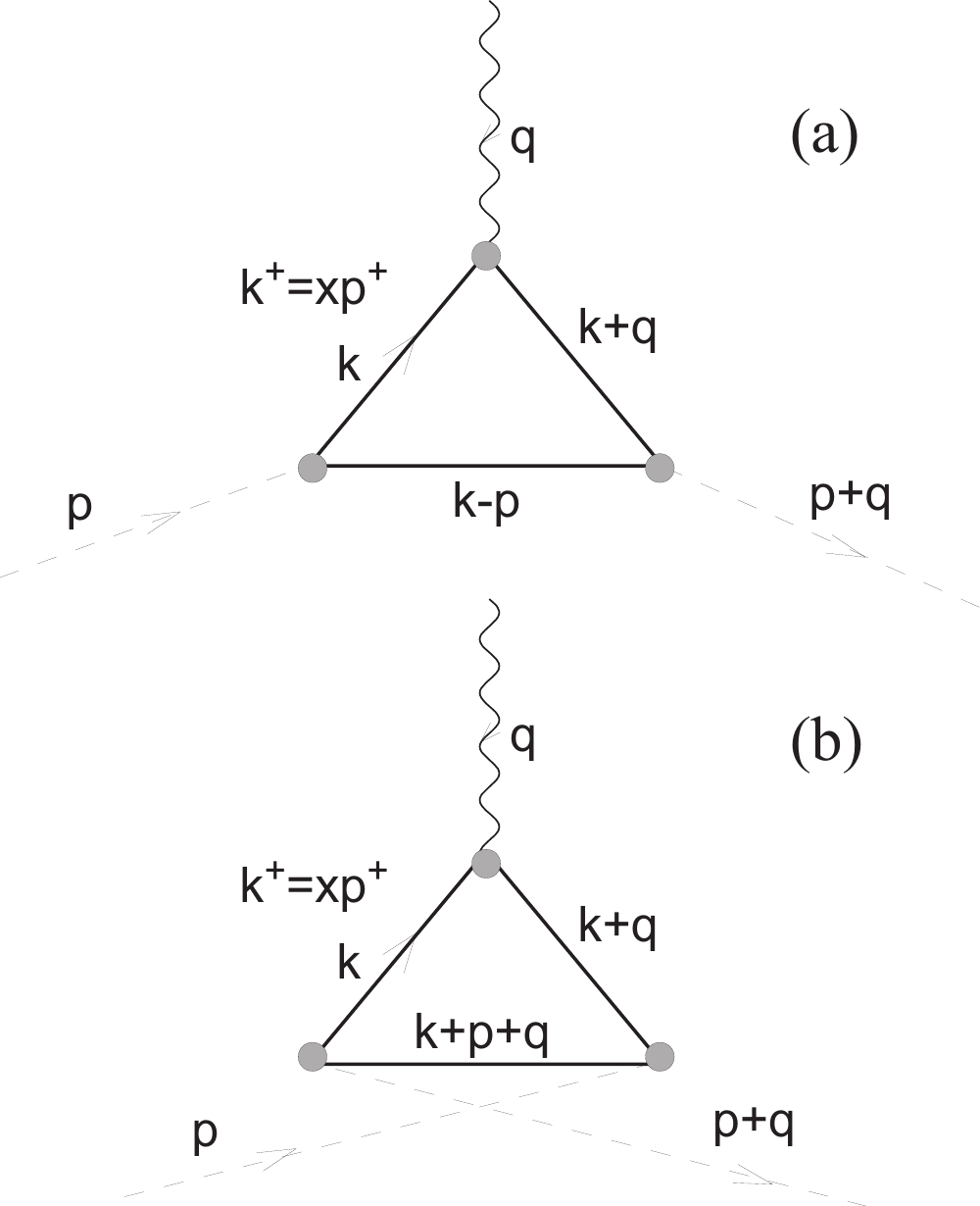} ~~ \includegraphics[angle=0,width=0.3 \textwidth]{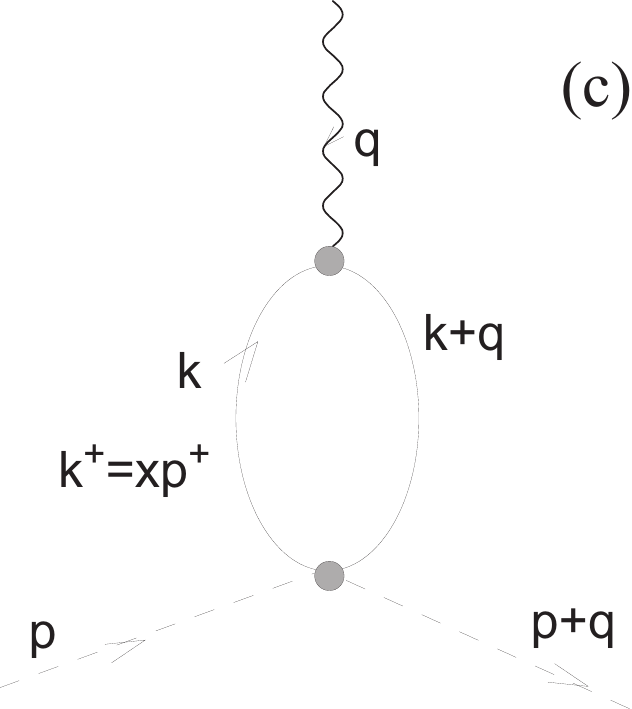}
\vspace{-1mm}
\caption{One-quark-loop diagrams for the evaluation of the quark GPDs of the pion at the quark model scale. \label{fig:diag}}
\end{figure}

Relations~(\ref{eq:t2ft3}) impose strict theoretic constraints on the
form of the off-shell GPDs via the moments~(\ref{eq:poly2}). It is
worthwhile to illustrate them in a nonperturbative model of the pion
GPDs. The dynamics of the pion, which is a pseudo-Goldstone boson of
the spontaneously broken chiral symmetry, can be effectively described
in (leading-$N_c$) chiral quark models. Particularly convenient for
our purpose in the spectral quark model (SQM) introduced
in~\cite{RuizArriola:2003bs}, where one overlays the contributions of
quarks of different (complex) mass, which serves as a regulator of the
high-energy contributions. In this model, complying to all the formal
requirements, one can exactly impose the vector meson dominance (VMD),
which is successful in the description of the charge form factor. The
evaluation proceeds according to the one-quark-loop diagrams of
Fig.~\ref{fig:diag}.

In the chiral limit of $m_\pi=0$ and in the half-off-shell case very simple expressions follow for the discussed form factors:
\begin{eqnarray}
&& \hspace{-7mm} F(t,p^2,0)= \frac{M_V^4}{\left(M_V^2-p^2\right) \left(M_V^2-t\right)}, \label{eq:sqmff}\nonumber\\
&& \hspace{-7mm}  G(t,p^2,0)=\frac{p^2 M_V^2}{\left(M_V^2-p^2\right) \left(M_V^2-t\right)}, \nonumber \\
&& \hspace{-7mm} \theta_1(t,p^2,0)=\frac{M_V^2 \left[\frac{p^2 (t-p^2)}{M_V^2-p^2}+(t-2 p^2) L \right]}{\left(t-p^2\right)^2}, \nonumber \\
&& \hspace{-7mm} \theta_2(t,p^2,0)   = \frac{M_V^2 \left[\frac{p^2 (p^2-t)}{M_V^2-p^2}+t L\right]}{\left(t-p^2\right)^2}, \nonumber \\
&& \hspace{-7mm} \theta_3(t,p^2,0)  = \frac{p^2 M_V^2 \left[p^2-t+(M_V^2-p^2)L \right]}{\left(t-p^2\right)^2
   \left(M_V^2-p^2\right)},
\end{eqnarray}
where $p^2$ is the off-shellness of one on the pions, $L=\log\frac{M_V^2-p^2}{M_V^2-t}$ and $M_V$ denotes the $\rho$
meson mass. The pion propagator in SQM has the form 
\begin{eqnarray}
\Delta(p^2)= \frac{M_V^2-p^2}{M_V^2 p^2}. \label{eq:psqm}
\end{eqnarray} 
We note that the charged form factors $F$ and  $G$ in Eqs.~(\ref{eq:sqmff})
exhibit factorization in $p^2$ and $t$, while this is not the case for the gravitational form factors $\theta_i$. 
Formulas~(\ref{eq:sqmff}) satisfy relations~(\ref{eq:t2ft3}).

\begin{figure}[t]
\centering
\includegraphics[angle=0,width=0.48 \textwidth]{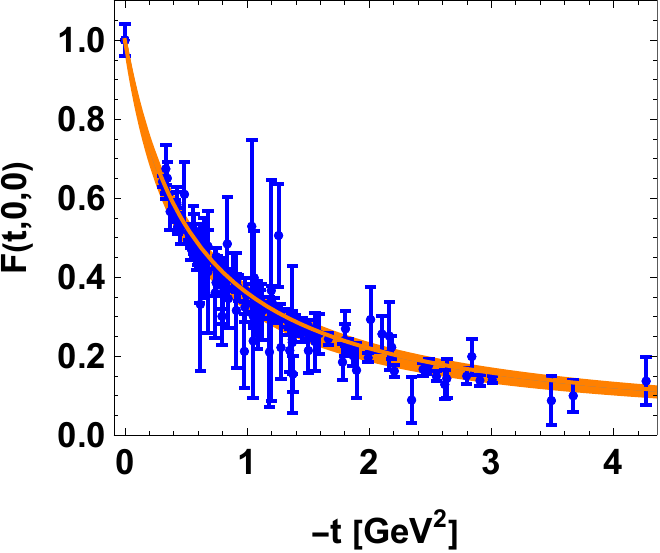} ~ \includegraphics[angle=0,width=0.49 \textwidth]{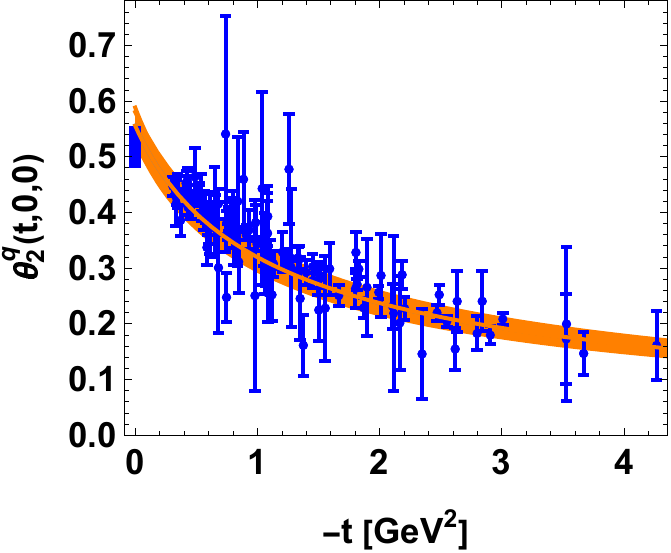} 
\vspace{-1mm}
\caption{Joint SQM fits to the lattice data~\cite{Brommel:2005ee,Brommel:PhD} for the on-shell $F$ and the quark part of $\theta_2$, yielding $M_V=0.75(5)$~GeV. The bands reflect the lattice errors (see~\cite{Broniowski:2008hx} for details). \label{fig:vgff}}
\end{figure}

For the on-shell gravitational form factors of the pion there exist
lattice data~\cite{Brommel:2005ee,Brommel:PhD}, which were used to fit
to the SQM formulas in~\cite{Broniowski:2008hx}. The result, together
with the charge form factor, is presented in Fig.~\ref{fig:vgff}. The
value of $\theta_2^q$ at $t=0$ indicates the momentum fraction carried
by the quarks at the lattice scale $\mu \sim 2$~GeV
(see~\cite{Broniowski:2008hx} for further discussion). Notably, the
distribution of matter inside the pion is more compact compared to the
distribution of charge, which in SQM is reflected by the following
relation between the corresponding means squared
radii~\cite{Broniowski:2008hx}:
\begin{eqnarray}
2 \langle r^2 \rangle_{\theta_2} =  \langle r^2 \rangle_F
\end{eqnarray}
which implies that matter is more concentrated than charge.\footnote{This chiral quark model
relation agrees with a simple vector and tensor meson dominance approach where
$2 \langle r^2 \rangle_{\theta_2} / \langle r^2 \rangle_F \sim (m_\rho/m_{f_2})^2
0.8(4)$ upon use of the half width rule~\cite{Masjuan:2012sk}}

\begin{figure}[t]
\centering
\includegraphics[angle=0,width=0.487 \textwidth]{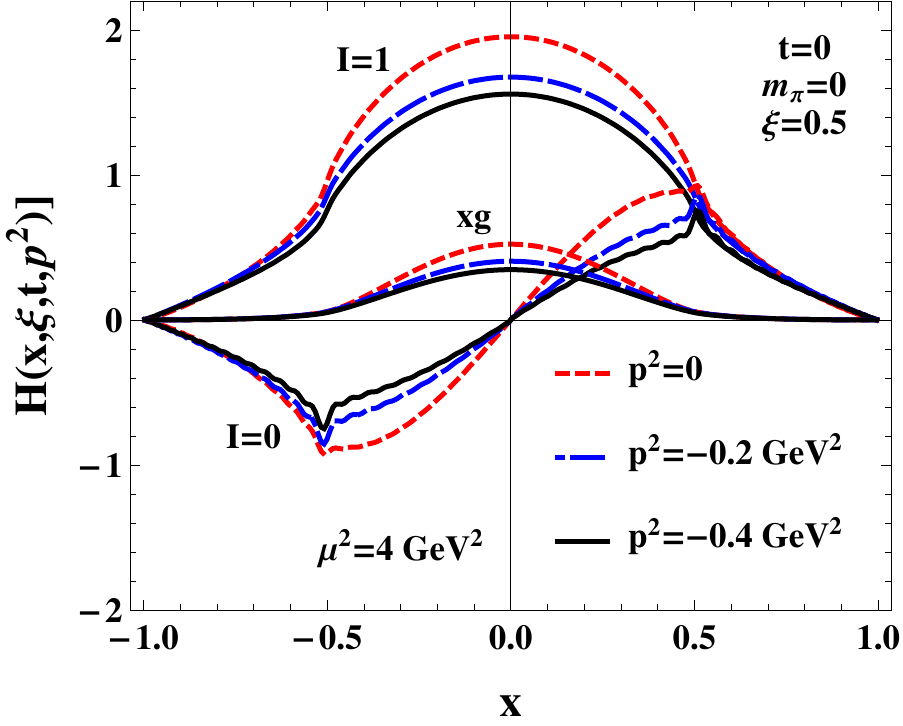} ~~ \includegraphics[angle=0,width=0.472 \textwidth]{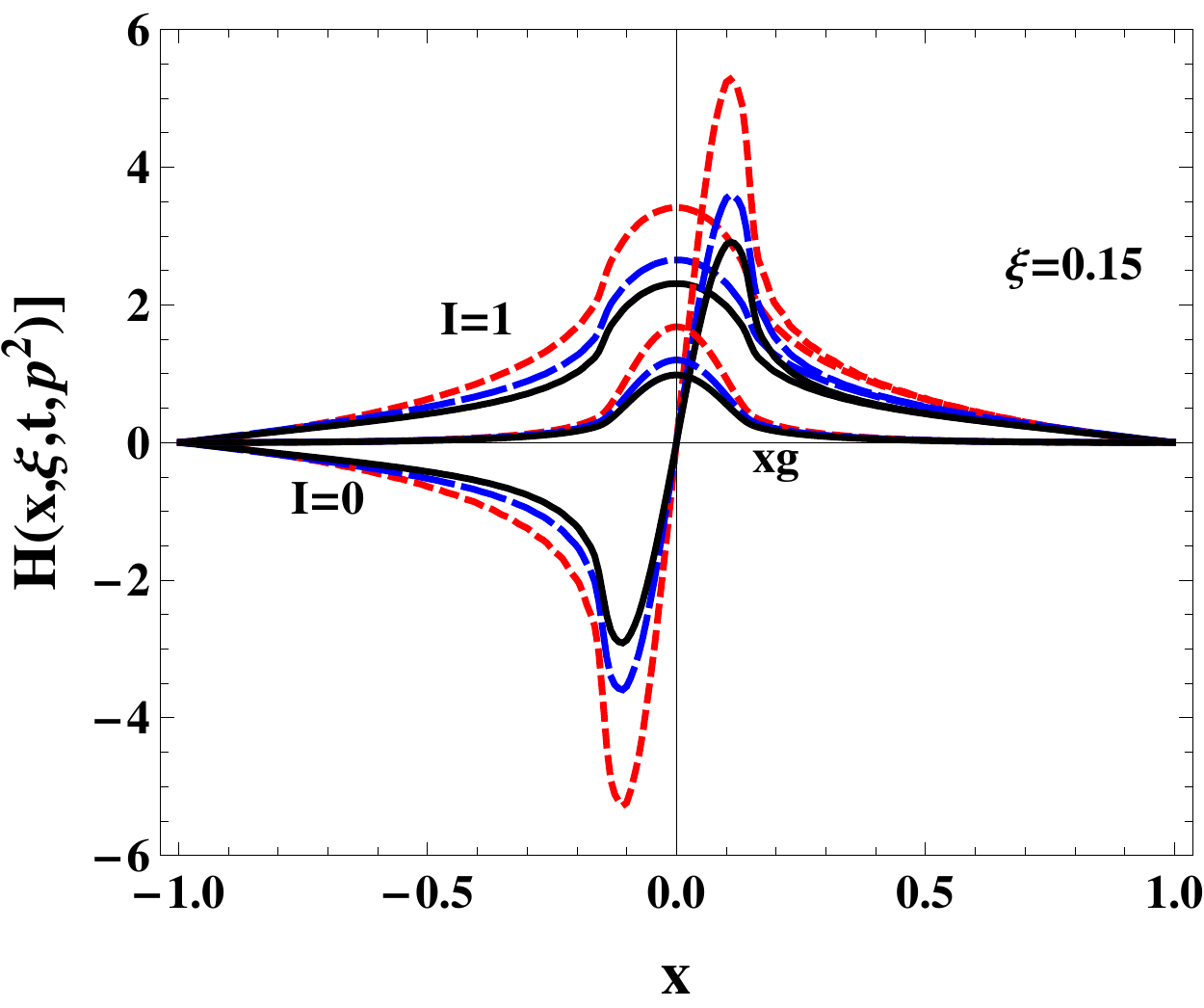} 
\vspace{-1mm}
\caption{Half-off-shell GPDs of the pion in SQM in the chiral limit at $t=0$, $\xi=0.5$ (left) and $\xi=0.15$ (right), evolved to $\mu=2~{\rm GeV}$. The three sets of lines correspond to the quark isovector GPD, quark isoscalar GPD, and the gluon GPD multiplied with $x$. \label{fig:gpd}}
\end{figure}

The half-off-shell GPDs in SQM can be evaluated analytically at the quark model
scale~\cite{Broniowski:2007si} $\mu_0$, where the valence quarks
carry $100\%$ of the pion's momentum. The formulas (which are lengthy, hence not shown) display a lack of factorization in $x$, $t$, and $p^2$.  
The GPDs are subsequently evolved from the scale $\mu_0$ to a
higher scale $\mu$ with the leading-order DGLAP-ERBL QCD evolution equations ~\cite{GolecBiernat:1998ja}. 
The result for $t=0$ and skewness values $\xi=0.5$ and $0.15$, evolved to $\mu = 2$~GeV, is presented in
Fig.~\ref{fig:gpd}. We notice a sizable dependence on the off-shellness $p^2$. At
the maxima of the curves at $p^2=-0.2~{\rm GeV}^2$ and $p^2=0$ we note a
relative effect of $\sim$10\% for the isovector GPD, and 
$\sim$20\% for the isoscalar GPDs. With $p^2=-0.4~{\rm GeV}^2$ the effects
are correspondingly higher, about 20\% and 35\%. 

\begin{figure}[t]
\centering
\includegraphics[angle=0,width=0.45 \textwidth]{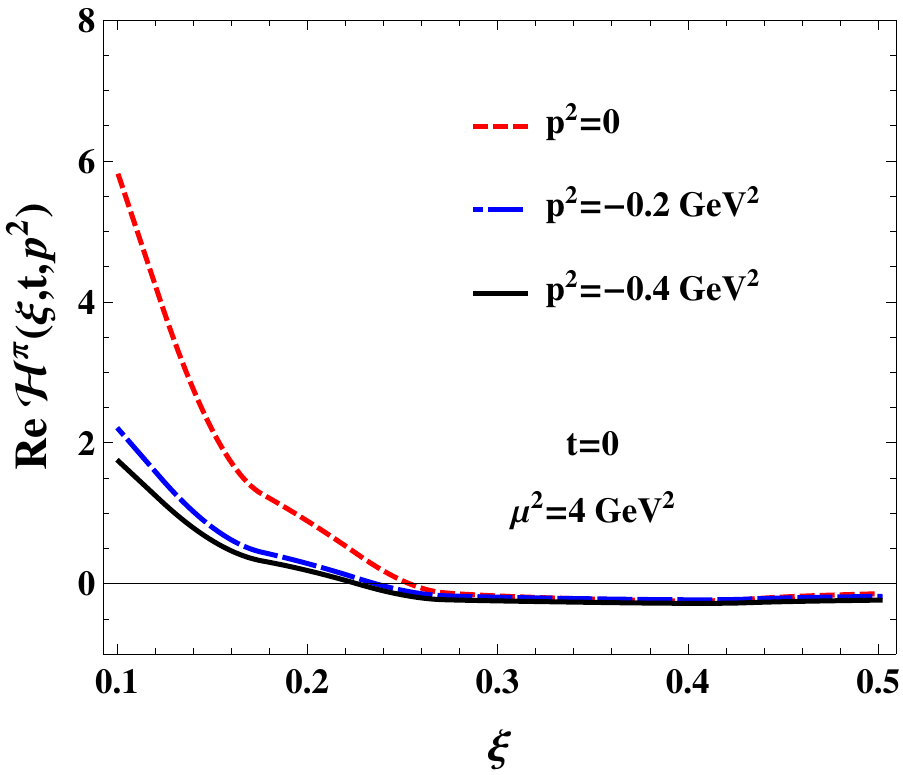} \includegraphics[angle=0,width=0.45 \textwidth]{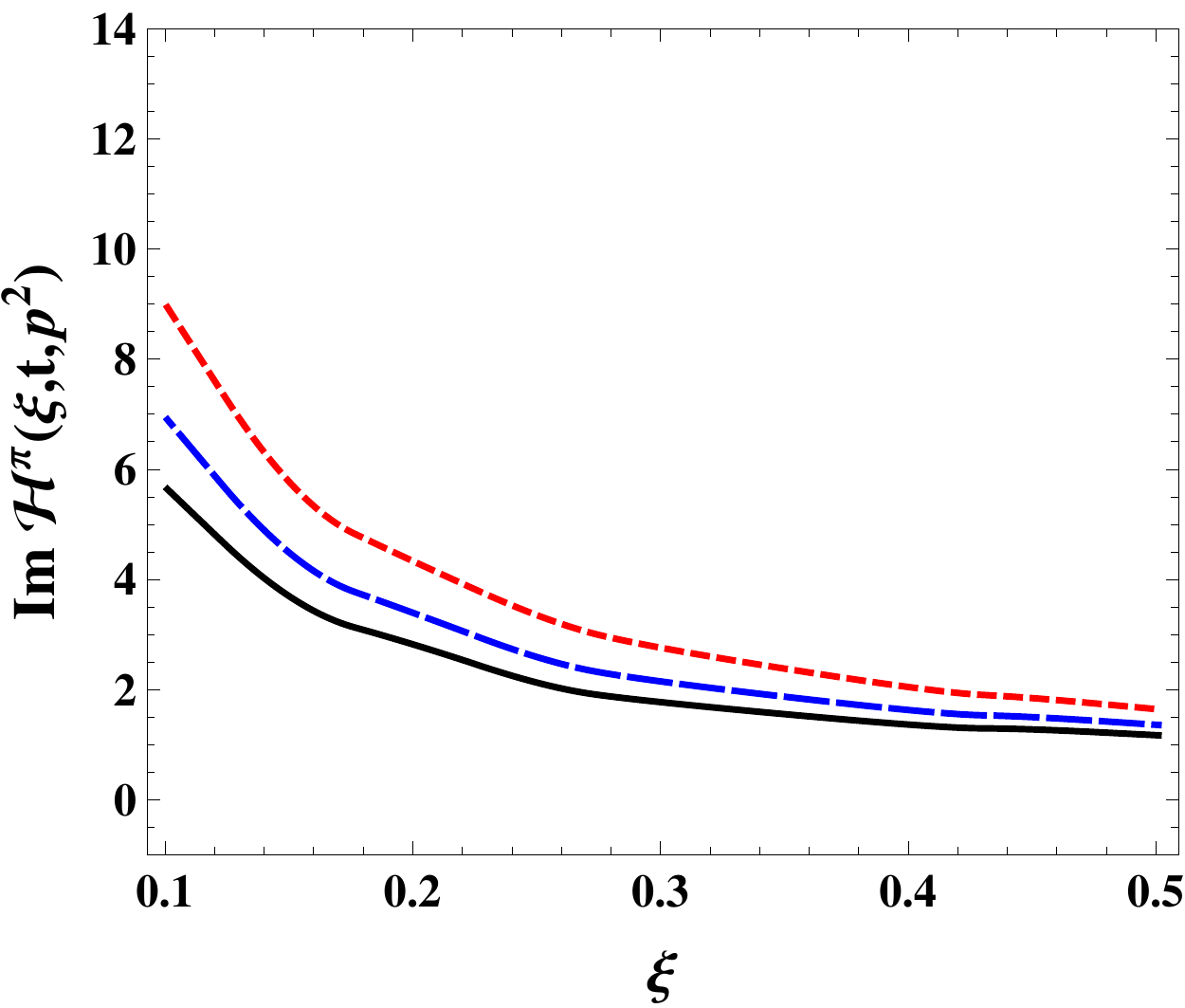} 
\vspace{-1mm}
\caption{Real and imaginary parts of the Compton amplitude from SQM, evolved to  $\mu=2~{\rm GeV}$.  \label{fig:comp}}
\end{figure}

What enters the evaluation of the amplitude of the Sullivan process is the Compton (DVCS) amplitude, defined as~\cite{Amrath:2008vx} 
\begin{eqnarray}
{\cal H}_{\pi^+}(\xi,t,p^2) = \sum_{q=u,\bar{d}} \!\! e_q^2 \int_{-1}^1 dx \left [  \frac{1}{\xi-x-i \epsilon} - \frac{1}{\xi+x-i \epsilon}   \right ] {H}_q(x,\xi,t,p^2), \label{eq:comp}
\end{eqnarray}
where we have used the half-off-shell kinematics. The evaluation of Eq.~(\ref{eq:comp}) with the GPDs of Fig.~\ref{fig:gpd} in the integrands yields the results shown in Fig.~\ref{fig:comp}. As discussed in~\cite{Amrath:2008vx},
${\cal H}_{\pi^+}$ interferes with the dominant Bethe-Heitler amplitude~\cite{Amrath:2008vx}, hence the uncertainties in
the GPDs are carried over linearly to the Sullivan cross section. One should remark here that NLO effects~\cite{Moutarde:2013qs} should be incorporated,  since then the gluons yield a relevant (dominating) contribution~\cite{MorgadoChavez:2022vzz}.

The off-shellness effects add up to other sources of model uncertainties in the estimates of
the GPDs, such as the value of the vector meson mass in SQM ($\sim$10-15\% in the form
factors~\cite{Masjuan:2012sk} and parton distributions), the use of the exact chiral limit (
$\sim$5\%~\cite{Davidson:1994uv}), or the uncertainty in the value of the
quark-model scale ($\sim$10\%~\cite{Broniowski:2007si}).

Finally, we present a general methodological point concerning the evaluation of amplitudes
such as for the Sullivan process in Fig.~\ref{fig:sul}, proceeding along the lines of~\cite{Koch:2001ii}. The off-shellness manifests itself
in all components of the Feynman diagram: in the GPDs/DVCS amplitudes, as discussed previously, but also in the pion propagator or the pion nucleon form factor (not covered in this talk). In particular, one should use the full pion off-shell propagator, and not just its pole approximation. If however, as is typically done phenomenologically, one includes only the pion pole term, $1/(p^2-m_\pi^2)$, instead of a full propagator $\Delta(p^2)$, one misses
the factor $\Delta(p^2) (p^2-m_\pi^2)=1/F(0,p^2,m_\pi^2)$. This factor can attributed to the half-off-shell vertex by defining
$\Gamma^{\ast \mu}(t,p^2,m_\pi^2)\equiv \Gamma^\mu(t,p^2,m_\pi^2)/F(0,p^2,m_\pi^2)$, which then should be used in calculations with the pion pole term, such as in the diagram in the right panel of Fig.~\ref{fig:sul}. Then for the charge form factor
\begin{eqnarray}
\Gamma^{\ast \mu}(t,p^2,m_\pi^2)&=&2P^\mu \frac{F(t,p^2,m_\pi^2)}{F(0,p^2,m_\pi^2)} -  q^\mu \frac{p^2}{t} \left[ 1 -\frac{F(t,p^2,m_\pi^2)}{F(0,p^2,m_\pi^2)} \right]. \label{eq:star}
\end{eqnarray}
When the dependence on $t$ and $p^2$ in $F$ factorizes, as in SQM in the chiral limit,  then
the dependence on $p^2$ remains only in the term proportional to $q^\mu$. Thus it is present for virtual photons,
while for a real photon it can be removed by a suitable choice of gauge~\cite{Koch:2001ii}. Similarly
to Eq.~(\ref{eq:star}), one could attribute the off-shell propagator correction $1/F(0,p^2,m_\pi^2)$ to the considered half-off-shell GPDs.

To conclude, we have analyzed off-shell effects in the GPDs of the pion and in the corresponding electromagnetic and gravitational
form factors. As the crossing symmetry is no longer effective, the polynomiality feature of GPDs involves also the odd powers of the skewness parameter $\xi$. However, WTIs result in relations between the off-shell charge and gravitational form factors, which may serve as important consistency constraints for the form of the off-shell GPDs. We have applied SQM to illustrate the general formalism, as well as to estimate the actual magnitude of the effects, after a suitable QCD evolution to the scale $\mu=2$~GeV. For the half-off-shell GPDs we find the relative (to the case of $p^2=0$) effect
at $p^2=-0.2~{\rm GeV}^2$ of $\sim$10\% for the isovector GPD, and 
$\sim$20\% for the isoscalar GPDs. At $p^2=-0.4~{\rm GeV}^2$ the effects
are correspondingly $\sim$20\% and $\sim$35\%. They carry over linearly to the DVCS amplitude, and thus to the cross section of the Sullivan process.

\medskip

We cordially thank Krzysztof Golec-Biernat for providing us with his GPD evolution code.
WB acknowledges the support by the Polish National Science Centre (NCN) grant 2018/31/B/ST2/01022, VS by NCN grant 2019/33/B/ST2/00613, and ERA by project PID2020-114767GB-I00 funded by MCIN/AEI/10.13039/\-501100011033, as well as Junta de Andaluc{\'i}a grant FQM-225.
 
\bibliography{Ref}

\end{document}